\documentclass[onecolumn,showpacs,preprintnumbers,amsmath,amssymb]{revtex4}
\usepackage{graphicx}
\usepackage{dcolumn}
\usepackage{bm}
\usepackage{color}
\usepackage{amsmath}
\usepackage{xcolor}
\usepackage{caption}
\usepackage{float}
\usepackage{soul}

\begin{document}

\title{Influence of coherent adiabatic excitation on femtosecond transient signals}

\author{A. Peralta Conde$^1$, R. Montero$^2$, and A. Longarte$^2$}

\affiliation{$^1$  Centro de L\'aseres Pulsados, CLPU, Parque Cient\'ifico, 37185 Villamayor, Salamanca, Spain. \\
$^2$ Departamento de Qu\'imica-F\'isica, Facultad de Ciencia
y Technology\'ia, Universidad del Pa\'is Vasco, Apartado 644, ES-48080
Bilbao, Spain}

\begin{abstract}
The transient signals derived from femtosecond pump-probe experiments are analyzed in terms of the
coherent evolution of the energy levels perturbed by the excitation pulse. The model system is treated
as the sum of independent two-level subsystems that evolve adiabatically or are permanently excited, depending on the detuning
from the central wavelength of the excitation laser. This approach will allow us to explain numerically and analytically the convergence between
the coherent and incoherent (rate equations) treatments for complex multi-level systems. It will be also shown that
the parameter that determines the validity of the incoherent
treatment is the distribution of states outside and inside the laser
bandwidth, rather than the density of states as it is commonly
accepted.
\end{abstract}


\maketitle

\section{Introduction}

The substantial evolution of laser sources during the last decades
has permitted to monitor, and even control processes taking place in
a broad range of time scales. More recently, the efforts in the
development of coherent sources of ultrashort pulse duration, i.e.,
attosecond (as) and femtosecond (fs), together with the increase
control over the laser pulses, have provided an ideal tool for
recording snapshots of ultrafast evolving systems (see for example
\cite{DRES02, UIBER07, ZEW88, CAVA07, HER96, GOR97, Skan10} and
reference therein). Nowadays, experiments with femtosecond
resolution in the fields of Chemistry, Physics and Biology are
carried out routinely by using a great variety of pump probe
methods, e.g., transient ionization, fluorescence up conversion,
fluorescence dip spectroscopy, and femtosecond time resolved Raman
\cite{ChemRev04}. In these techniques, the system is excited by one
(or more) photons of the pump pulse triggering simultaneously in all
the particles the process of interest. After a variable temporal
delay, the absorption of another (or more) photons from the probe
laser produces a measurable signal. The dependence of the system
response on the pump probe delay contains the desired information on
the dynamics of the process.

Modeling the collected time dependent signals is pivotal for the
correct extraction of the information contained in the experimental
data. There are two possible approaches for this task: an incoherent
treatment based on rate equations, and a coherent model where the
interaction of the system with the pump pulse is described by the
solution of the time dependent Schr\"{o}dinger equation (or the
equivalent Bloch equations). Let us consider that the system can be
described by a two-state model. Three different regimes can be
distinguished. If the losses dominates the dynamics of the system,
the coherence induced by the laser is lost very rapidly, i.e., we
can apply adiabatic elimination \cite{Sh90}, and therefore the
incoherent treatment provides an accurate description of the
dynamics. On the other hand, if the losses are much slower than the
pump pulse duration, the incoherent treatment can be applied once
again if one is not interested in reproducing the excitation
process. Finally, in a situation where the dynamics of interest is
of the order of the excitation pulse, the temporal information is
given by the solution of the time dependent Schr\"{o}dinger
equation.

However, this simple two-state model is in principle not applicable
when the pump laser bandwidth is large enough to couple more than
one single state, as it is usually found in molecules, clusters, and
solids. In this scenario, a full theoretical treatment requires of
very expensive quantum dynamical calculations, and it is usually not
accessible, being the ordinary approach to reduce this coherently
excited multistate system to an incoherently excited two-state
system. The accuracy of this approximation has been tested elsewhere
\cite{IHER06}, proving to be very effective as long as the
determination of the zero delay between pump and probe laser pulses
is made correctly.

The aim of the present paper is to explore the applicability of the coherent and incoherent approximation to the
modeling of femtosecond transients derived from molecular dynamics. Although the coherent excitation of quasicontinuum systems
has been a matter of intense research, and the effect of dephasing and density of states well documented experimentally,
and rigorously described theoretically (see for example \cite{IHER06, MUK95, KYR85} and references therein), herein we adopt a different
simpler approach from the perspective of the adiabatic evolution of the levels outside the laser bandwidth. This phenomenon is generally known as
Coherent Population Return (CPR) \cite{Vi01, Al06}.  In this model a coherently excited multistate system is treated as the sum of independent coherently excited
two-state systems with variable detuning from the central wavelength
of the excitation laser. This procedure will allow us to justify
qualitatively, and analytically with the help of Fourier analysis,
the use of the incoherent approximation for such complex systems. We
will provide analytical expressions to quantify the error carried
out by this treatment, demonstrating that the critical parameter
that determines the validity of this approach is the distribution of
states with respect to the laser bandwidth, rather than the density
of states as it is commonly accepted. Furthermore, this treatment
underlines the importance of considering CPR for a correct
description of laser-matter interaction \cite{Al10, Mon10}

\section{Theoretical models}

\subsection{Two-state system}

For coherent interactions the description of the population dynamics
is provided by the time dependent Schr\"{o}dinger equation
\begin{equation}
\label{timedepsch} i\hbar\frac{\partial\Psi(t)}{\partial
t}=H(t)\Psi(t),
\end{equation}
where $\Psi(t)$ is the statevector of the system, and $H(t)$ the
Hamiltonian including the interaction with a radiation field.

\begin{figure}[ht!]
\begin{center}
\includegraphics[width=4cm, height=3.5cm]{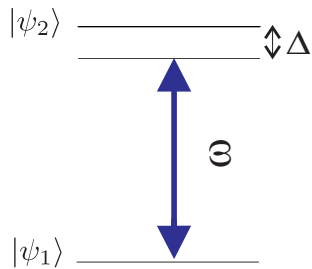}
\caption{\label{twolevelsytem} Two-state system interacting with a
radiation field. The laser detuning from resonance is represented by
$\Delta$. The energies of the excited and ground states are $E_2$
and $E_1$ respectively.}
\end{center}
\end{figure}

Let us consider a two-state system interacting with the electric
field of a laser pulse with carrier frequency $\omega$, detuned from
the transition frequency by $\Delta$. This situation is described by
Fig\,\ref{twolevelsytem}.

The time dependent Hamiltonian in the electric dipole approximation
can be written as
\begin{equation}
\label{hamidefi} H(t)=H^0+H^I(t),
\end{equation}
where $H^0$ is the two-state field free Hamiltonian with energies
$E_1$ and $E_2$
\begin{equation}
H^0= \left[
\begin{array}{cc}
E_1 & 0 \\
0 & E_2
\end{array}
\right],
\end{equation}
and $H^I$ is the interacting Hamiltonian
\begin{equation}
H^I(t)= \left[
\begin{array}{cc}
0  & \langle \psi_1
|\textrm{V}| \psi_2 \rangle   \\
\langle \psi_2 |\textrm{V}| \psi_1 \rangle   & 0
\end{array}
\right],
\end{equation}
with
\begin{equation}
\textrm{V}=-\textbf{d}\cdot[\textbf{e}\mathcal{E}(t)\cos \omega t],
\end{equation}
being $\textbf{d}$ the dipole moment, and $\textbf{e}$ and
$\mathcal{E}(t)$ the polarization direction and the slow varying
envelope of the electric field respectively.

If as customary the rotating wave approximation (RWA) is applied for
neglecting the fast oscillating terms \cite{Sh90}, the
RWA-Hamiltonian reads
\begin{equation}
\label{ch2hamiltonian}
 H^{RWA}=\frac{\hbar}{2}\left[
\begin{array}{cc}
0 & \Omega(t)\\
\Omega(t) & 2\Delta
\end{array}
\right],
\end{equation}
where $\Omega(t)$ is the Rabi frequency, i.e., the interaction
energy divided by $\hbar$,
\begin{equation}
\label{ch2definitonofrabi}
\Omega(t)=-(\textbf{d}\cdot\textbf{e})\mathcal{E}(t)/\hbar,
\end{equation}
and $\Delta$ the laser detuning
\begin{equation}
\Delta=\frac{(E_2-E_1)}{\hbar}-\omega.
\end{equation}

It is convenient to describe the population dynamics in the
adiabatic basis $\left[\Phi_+, \Phi_-\right]$,which is the basis
formed by the instantaneous eigenstates of the Hamiltonian described
by Eq.\,\ref{ch2hamiltonian}. These adiabatic eigenstates can be
written as
\begin{equation}
\label{Phi+} \Phi_+ (t)=\psi_1\sin \vartheta(t)+\psi_2 \cos
\vartheta(t),
\end{equation}
\begin{equation}
\label{Phiminus} \Phi_- (t)=\psi_1\cos \vartheta(t)-\psi_2 \sin
\vartheta(t),
\end{equation}
where $\vartheta(t)$ is the mixing angle
\begin{equation}
\label{mixingangle}
\vartheta(t)=\frac{1}{2}\arctan\frac{\Omega(t)}{\Delta},
\end{equation}
and $\psi_j$ ($j=1,2$) are the bare states. The corresponding
eigenenergies are
\begin{equation}
\lambda_\pm=\frac{1}{2}\left[\Delta\pm\sqrt{\Omega^2(t)+\Delta^2}\right].
\end{equation}

\begin{figure}
\includegraphics[width=7cm, height=6cm]{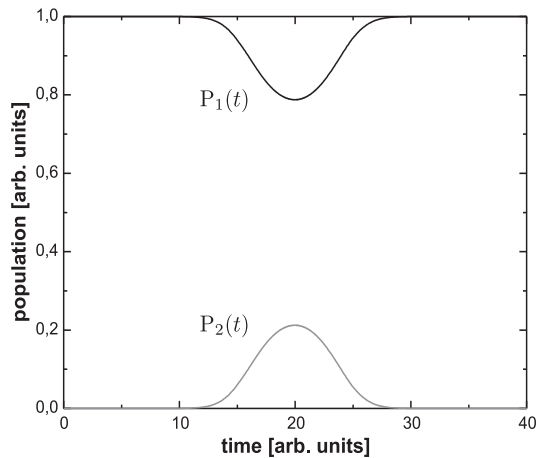}
\caption{\label{popubare} Population dynamics in the case of
Coherent Population Return (see Eq.\,\ref{Popu} for the analytical
expressions of the populations). P$_i$ is the population of state i
(i=1, 2).}
\end{figure}

Let us assume $\Omega(t)$ to be negligibly small outside a finite
time interval $t_i < t < t_f$, i.e., outside the pulse duration
$\tau = t_f-t_i$. Consider now the case of the laser frequency
detuned from exact resonance, i.e., $\Delta\gtrsim1/\tau$. If at the
beginning of the interaction all the population is in the ground
state, the statevector of the system $\Psi(t)$ at time $t=-\infty$
is aligned parallel to the adiabatic state $\Phi_-(t)$ because
$\Psi(-\infty)=\Phi_-(-\infty)=\psi_1$. If the evolution of the
system is adiabatic, that means the Hamiltonian varies sufficiently
slow in time, the statevector of the system $\Psi(t)$ remains always
aligned with the adiabatic state $\Phi_-(t)$. Thus, during the
excitation process when $t_i<t<t_f$ and $\Omega(t)\neq 0$, the
statevector $\Psi(t)$ is a coherent superposition of the bare states
(see Eq.\,\ref{Phiminus}). Therefore, some population is transiently
excited to the upper state. However at the end of the interaction
when $t=+\infty$, the statevector of the system becomes once again
aligned with the initial state $\psi_1$, i.e.,
$\Psi(+\infty)=\Phi_-(+\infty)=\psi_1$. The population transferred
during the process from the ground state to the excited state
returns completely to the ground state after the excitation process.
No population resides permanently in the excited state, no matter
how large the transient intensity of the laser pulse may be (see
Fig.\,\ref{popubare}). This coherent phenomenon is called Coherent
Population Return (CPR) \cite{Vi01, Al06}. It can be shown
\cite{Vi01} that the condition for smooth adiabatic evolution -- and
hence CPR-- is just the adiabatic condition $|\Delta|\gtrsim1/\tau$.
It is important to notice that the adiabatic condition is
independent of $\Omega(t)$, and therefore of the laser intensity.

It is necessary to consider now a probe process to detect the
population transferred to state $\psi_2$, e.g., by ionization of the
excited state with a short probe laser pulse. According to our
previous discussion, if the systems evolves adiabatically
($|\Delta|\gtrsim1/\tau$) no population remains in the excited state
once the pump pulse is over due to CPR, and the ionization signal is
produced exclusively when pump and probe pulses overlap in time. On
the other hand, if the evolution of the system is diabatic, meaning
$|\Delta|\lesssim1/\tau$, there is a permanent population
transferred to the excited state, and consequently, ionization
signal when pump and probe pulse do not temporally overlap.

\subsection{Extension of the two-state model to multi-state systems}

\begin{figure}[ht!]
\begin{center}
\includegraphics[width=8cm, height=5.21cm]{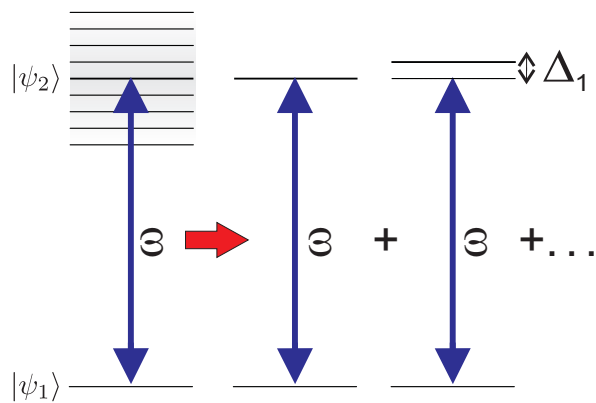}
\caption{\label{twolevelsytem_multiple} Multistate system as the sum
of independent two-state systems with different detunings.}
\end{center}
\end{figure}

So far, we have assumed a two-state system. However, this simple
model can not be usually applied to real systems where inside the
laser bandwidth a substantial number of states are excited, being
necessary to consider a more general situation. If as a first
approximation we neglect any interaction between the excited states,
and assume that the losses of the system are much slower than the
excitation process, we can model a multistate system by means of the
sum of multiple individual two-state systems with different
detunings with respect to the excitation laser (see
Fig.\,\ref{twolevelsytem_multiple}). It is worth to notice that the
mechanisms that couple different excited levels as internal
vibrational redistribution (IVR) in the case of vibronic levels,
take place usually on hundreds of femtoseconds. Thus, for fs laser
pulses this approximation is valid.\\

\begin{figure}
\includegraphics[width=8cm, height=5.6cm]{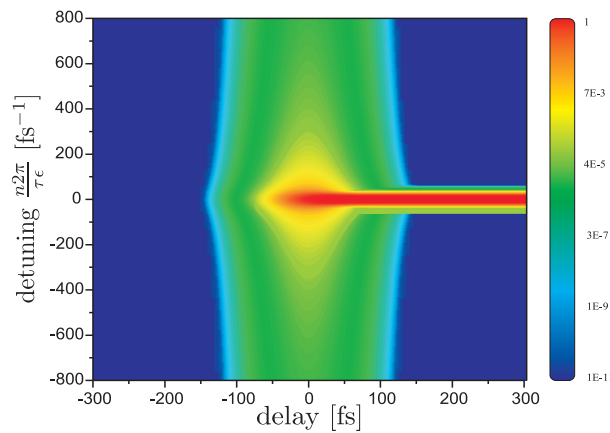}
\caption{\label{detunings} Ionization signal versus detuning and
delay between pump and probe laser (1+3').  The parameters chosen
for the simulation were: pump pulse duration (full width at half
maximum (FWHM) of the peak intensity) $\tau=50$ fs, probe pulse
duration $\tau_{Pr}=50$ fs, peak Rabi frequency
$\Omega=9.4\,10^{-4}\,\textrm{fs}^{-1}$, peak ionization loss rate
$\Gamma=4\,10^{-6}\,\textrm{fs}^{-1}$, number of states per
$1/\tau$\, $\epsilon=50$.}
\end{figure}

\begin{figure}
\includegraphics[width=8cm, height=5.6cm]{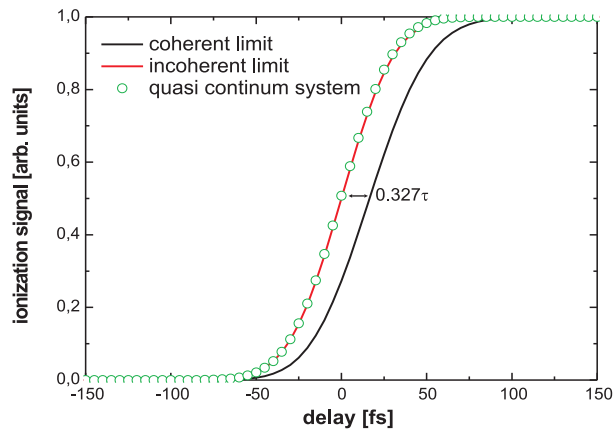}
\caption{\label{comparison} Normalized ionization signal as a
function of the delay between pump and probe pulses, for the
different possible approaches to the problem: solution of the time
dependent Schr\"{o}dinger equation for a two-state system on
resonance (black solid line), solution of rate equations (red solid
line), and multiple two-state approximation (green circles). $\tau$
is the pulse duration full width at half maximum (FWHM) of the peak
intensity.}
\end{figure}

Figure\,\ref{detunings} shows the ionization signal versus detuning
of the pump pulse, and temporal delay between pump and probe pulses,
for the system proposed in Fig.\,\ref{twolevelsytem}. As we have
previously discussed, we can see that only for those detunings that
do not fulfilled the adiabatic condition ($|\Delta|\lesssim
1/\tau$), there is  a permanent transfer of population to the
excited state, and therefore, a transient ionization signal when
pump and probe laser pulses do not temporally overlap. According to
the model described in Fig.\,\ref{twolevelsytem_multiple}, the total
ionization signal will be the sum of the ionization signals obtained
for different $\Delta$. This is shown in Fig.\,\ref{comparison}
together with the solution to the time dependent Schr\"{o}dinger
equation on resonance (coherent limit), and the solution of the
kinetic model (incoherent limit) for a two-state system. We can see
that the coherent excitation of a multistate system, under certain
conditions that will be discussed in the next section, is equivalent
to the incoherent excitation of a two-state system. This well known
result \cite{IHER06} simplifies enormously the analysis of pump
probe experiments involving a large number of states. Additionally,
for the reliable interpretation of the experimental data, it is
necessary an exact determination of the zero delay between pump and
probe laser pulses. A widely used  method to determine the zero
delay is to assign it to the position where the excitation signal of
a long living state reaches half of its maximum value.
Figure\,\ref{comparison} shows that this criterion is only valid in
the case of the incoherent limit, while for the coherent treatment
the 1/2 value is shifted $0.327\tau$ respect to the zero delay
position. Therefore, in establishing the zero delay time is
necessary to know beforehand what kind of behavior the problem
system presents.

\subsection{Results and discussion}

\subsubsection{Qualitative convergence of the coherent and incoherent models} \label{qualitative}

\begin{figure}
\includegraphics[width=8cm, height=5.6cm]{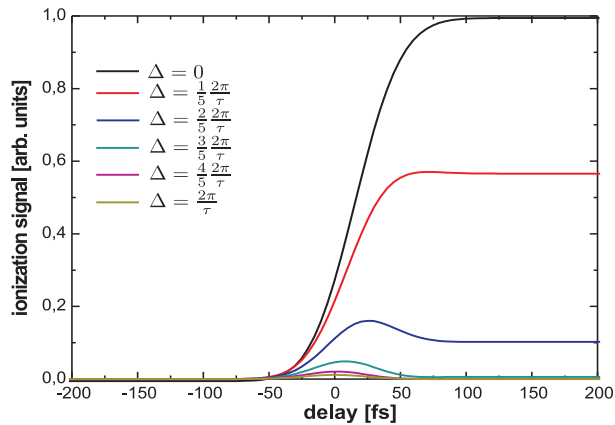}
\caption{\label{diff_detunings}  Ionization signal versus delay
between pump and probe pulses for different detunings. All the
signals are normalized with respect to the maximum signal obtained
at $\Delta=0$.}
\end{figure}

According to our previous discussion, the ionization signal of a
quasi continuum system has two different contributions: a step-like
transient signal for those detunings that do not fulfilled the
adiabatic condition, and a symmetric signal with respect to the
delay between pump and probe laser pulses, for those detunings that
fulfilled the adiabatic condition (see Fig.\,\ref{diff_detunings}).
When both contributions are added up, the latter is the responsible
for the change in the slope that shifts the transient signal to
earlier times (see Fig.\,\ref{comparison}). Since the contributions
to the signal coming from the non adiabatic region
($\Delta\lesssim1/\tau$), is much more intense than the one coming
from the adiabatic region ($\Delta\gtrsim 1/\tau$), it is necessary
to have a large number of states inside the adiabatic region, i.e.,
outside the pump laser bandwidth, to reach the convergence to the
incoherent limit. If this is not the case, the final situation will
be an intermediate case between the coherent and the incoherent
limits, being necessary a more elaborated analysis of the
experimental results.

\subsubsection{Analytical demonstration}

The convergence of the incoherent and the coherent treatments can
also be demonstrated analytically. Let us define the intensity like

\begin{equation}
\label{Intensity_1} \textrm{I}(t)=\textrm{I}_0 f(t)^2=\textrm{I}_0
e^{-4\textrm{Ln}2\frac{t^2}{\tau ^2}}
\end{equation}

where $\textrm{I}_0$ is the peak intensity, and $\tau$ the pulse
duration (FWHM), the Rabi frequency like

\begin{equation}
\label{Rabi_1}
 \Omega(t)=\Omega_0 f(t)=\Omega_0
e^{-2\textrm{Ln}2\frac{t^2}{\tau ^2}}
\end{equation}

where $\Omega_0$ is the peak Rabi frequency, and let us suppose a
two-state system (see Fig.\,\ref{twolevelsytem}). In the case of
incoherent excitation, it can be easily shown that the evolution of
the population in the target state is described by

\begin{equation}
\textrm{P}_2(t)=\frac{1}{2}\left(1-\textrm{exp}\left(-\alpha\int_{-\infty}^t
\textrm{I}(t^{\prime})\textrm{d}t^{\prime}\right)\right)
\end{equation}
where $\alpha$ is the absorption coefficient \cite{Sh90}.\\

Considering a situation where the population of the ground state is
barely perturbed, which is a reasonable approximation for fs pulses
excitation, the population in the excited state reads:
\begin{equation}
\textrm{P}_2(t)=\frac{\alpha}{2}\int_{-\infty}^t
\textrm{I}(t^{\prime})\textrm{d}t^{\prime}.
\end{equation}

For coherent excitation the description of the population evolution
is based on the solution of the time dependent Schr\"{o}dinger
equation with an appropriate Hamiltonian. Taking into account the
time dependent Hamiltonian in the electric dipole approximation for
a two-state system described by Eq.\,\ref{ch2hamiltonian}, the time
dependent Schr\"{o}dinger equation reads

\begin{equation}
\label{Schr-equation} \imath \hbar\frac{\partial}{\partial t}\left(
\begin{array}{cc}
C_1(t) \\
C_2(t)
\end{array}
\right)=(\hbar/2)\left[
\begin{array}{cc}
0 & \Omega(t) \\
\Omega(t) & 2\Delta
\end{array}
\right]\left(
\begin{array}{cc}
C_1(t) \\
C_2(t)
\end{array}
\right),
\end{equation}
being $C_j(t)$ the probability amplitude of state j (j=1, 2) at time
t.\\
If, as mentioned before, the ground state is barely perturbed, i.e.,
$C_1(t)\simeq1$, the differential equation that describes the
temporal evolution of $C_2(t)$ is given by
\begin{equation}
\label{diffeC2} \frac{d}{dt}C_2(t)+\imath\Delta C_2(t)+
(\imath/2)\Omega(t)=0.
\end{equation}
Hence, the temporal evolution of $C_2(t)$ can be written as

\begin{equation}
C_2(t)=-\frac{\imath}{2} e^{-\imath\Delta t}\left(\int_{-\infty}^t
e^{\imath\Delta t^\prime} \Omega(t^\prime)dt^\prime\right),
\end{equation}

and the population in the excited state is given by
\begin{eqnarray}
\label{Popu} \textrm{P}_2(t)=|C_2(t)|^2=\left[\int_{-\infty}^t
\frac{\Omega(t^\prime)}{2} \cos (\Delta t^\prime) dt^\prime\right]^2
\nonumber \\+\left[\int_{-\infty}^t \frac{\Omega(t^\prime)}{2} \sin
(\Delta t^\prime) dt^\prime\right]^2.
\end{eqnarray}

Once we have obtained the population evolution for coherent and
incoherent excitation, and taking into account the model described
by Fig.\,\ref{twolevelsytem_multiple}, and the numerical results
presented in Fig.\,\ref{comparison}, we obtain the following
equality

\begin{widetext}
\begin{equation}
\label{equality} \frac{\frac{\alpha}{2}\int_{-\infty}^t
\textrm{I}(t^{\prime})\textrm{d}t^{\prime}}
{\frac{\alpha}{2}\int_{-\infty}^{+\infty}
\textrm{I}(t^{\prime})\textrm{d}t^{\prime}}=
\frac{\sum_{n=-\infty}^{+\infty}\left[ \left[\int_{-\infty}^t
\frac{\Omega(t^\prime)}{2} \cos (\frac{n 2 \pi}{ \tau \epsilon}
t^\prime) dt^\prime\right]^2+\left[\int_{-\infty}^t
\frac{\Omega(t^\prime)}{2} \sin (\frac{n 2 \pi}{ \tau \epsilon}
t^\prime)
dt^\prime\right]^2\right]}{\sum_{n=-\infty}^{+\infty}\left[\left[\int_{-\infty}^{+\infty}
\frac{\Omega(t^\prime)}{2} \cos (\frac{n 2 \pi}{ \tau \epsilon}
t^\prime) dt^\prime\right]^2+\left[\int_{-\infty}^{+\infty}
\frac{\Omega(t^\prime)}{2} \sin (\frac{n 2 \pi}{ \tau \epsilon}
t^\prime) dt^\prime\right]^2\right]},
\end{equation}
\end{widetext}

where we have introduced a normalization factor, and have defined
the detuning as $\Delta=\frac{n 2 \pi}{ \tau \epsilon}$, being $n$
an integer that takes into account different states as showed in
Fig.\,\ref{twolevelsytem_multiple}, and $\epsilon$ the number of
states contained in $1/\tau$. It is worth mentioning that the factor
$2\pi$ has been included to express $\Delta$ in terms of angular
frequency.

With the help of the derived expressions, we can easily justify the
factor 0.327$\tau$ that differentiates the temporal evolution of
P$_2(t)$ in the coherent and incoherent limits (see
Fig.\,\ref{comparison}). For incoherent excitation is
straightforward to check that the value P$_2(t)=\frac{1}{2}$ is
obtained at $t=0$. In the case of coherent excitation we have,

\begin{equation}
\label{factor12_coher} \frac{\left[\int_{-\infty}^t
\frac{\Omega(t^\prime)}{2}
dt^\prime\right]^2}{\left[\int_{-\infty}^{+\infty}\frac{\Omega(t^\prime)}{2}
dt^\prime\right]^2}=\frac{1}{2}.
\end{equation}

Substituting Eq.\,\ref{Rabi_1} into Eq.\,\ref{factor12_coher} we
obtain

\begin{equation}
\label{factor12_b}
 \frac{1}{2} \left(1+\textrm{Erf} \left[\frac{t\sqrt{2\ln
 2}}{\tau}\right]\right)=\frac{\sqrt{2}}{2}.
\end{equation}

where Erf is the error function

\begin{equation}
\textrm{Erf}(x)=\frac{2}{\sqrt{\pi}}\int_0^x e^{-z^2}dz.
\end{equation}

From Eq.\,\ref{factor12_b}

\begin{equation}
\textrm{Erf} \left[\frac{t\sqrt{2\ln
 2}}{\tau}\right]=\sqrt{2}-1,
\end{equation}
and therefore

\begin{equation}
\frac{t\sqrt{2\ln
 2}}{\tau}=0.385,
\end{equation}
obtaining finally

\begin{equation}
t=\frac{0.385\, \tau}{\sqrt{2\ln 2}}\simeq0.327\, \tau.
\end{equation}

In the next section we will apply Fourier analysis concepts to
derive an analytical justification of Eq.\,\ref{equality}.

\subsubsection{Fourier analysis}

Let us consider a periodic function $f(t)$ with period $a$ and its
Fourier series expansion
\begin{equation}
S_N=\frac{1}{2}A_0+\sum_{n=1}^{N}\left[A_n\cos\left(\frac{n\pi
t}{a}\right)+B_n\sin\left(\frac{n\pi t}{a}\right)\right]
\end{equation}
where the coefficients $A_0$,$A_n$, and $B_n$ are

\begin{eqnarray}
\label{Fourier_coeffic}
 A_0=\frac{1}{a}\int_{-a}^{a}f(t)dt \nonumber\\ A_n=\frac{1}{a}\int_{-a}^{a}f(t)\cos\left(\frac{n\pi
t}{a}\right)dt \nonumber\\
B_n=\frac{1}{a}\int_{-a}^{a}f(t)\sin\left(\frac{n\pi t}{a}\right)dt.
\end{eqnarray}

Thus we have

\begin{equation}
f(t)\sim S_N.
\end{equation}

It can be shown \cite{Walker88} that

\begin{eqnarray}
\label{Parseval_1} \int_{-a}^{a}\left(f(t)-S_N\right)^2dt= \nonumber
\\ \left(\frac{1}{a}\int_{-a}^{a}f(t)^2dt-\left(\frac{1}{2}A_0^2+\sum_{n=1}^N\left(A_n^2+B_n^2\right)\right)\right).
\end{eqnarray}

If we let $N$ tend to infinity, the convergence of the Fourier
series $S_N$ to $f(t)$ implies that the left side in
Eq.\,\ref{Parseval_1} is equal to zero, and therefore we obtain

\begin{equation}
\label{Parseval_2}
\frac{1}{a}\int_{-a}^{a}f(t)^2dt=\frac{1}{2}A_0^2+\sum_{n=1}^\infty\left(A_n^2+B_n^2\right),
\end{equation}
that is known as the Parseval's equality.\\

Let us define now the following coefficients:

\begin{eqnarray}
\label{Fourier_coeffic_modif}
 A_{0t}=\frac{1}{a}\int_{-a}^{t}f(t')dt'\nonumber\\ A_{nt}=\frac{1}{a}\int_{-a}^{t}f(t')\cos\left(\frac{n\pi
t'}{a}\right)dt' \nonumber\\
B_{nt}=\frac{1}{a}\int_{-a}^{t}f(t')\sin\left(\frac{n\pi
t'}{a}\right)dt'.
\end{eqnarray}

Substituting the above coefficient in the Parseval's equality we can
write

\begin{eqnarray}
\frac{\frac{1}{a}\int_{-a}^{t}f(t')^2dt'+\frac{1}{a}\int_{t}^{a}f(t')^2dt'}{\frac{1}{a}\int_{-a}^{a}f(t)^2dt}=
\nonumber \\ \frac{\frac{1}{2}
\left(A_{0t}+\left(A_0-A_{0t}\right)\right)^2}
{\frac{1}{2}A_0^2+\sum_{n=1}^\infty\left(A_n^2+B_n^2\right)}+ \nonumber \\
\frac{\sum_{n=1}^\infty\left(\left(A_{nt}+\left(A_n-A_{nt}\right)\right)^2+\left(B_{nt}+\left(B_n-B_{nt}\right)\right)^2\right)}
{\frac{1}{2}A_0^2+\sum_{n=1}^\infty\left(A_n^2+B_n^2\right)}.
\end{eqnarray}

Rearranging we obtain the following expression
\begin{equation}
\label{Parseval_3}
\frac{\frac{1}{a}\int_{-a}^{t}f(t')^2dt'}{\frac{1}{a}\int_{-a}^{a}f(t)^2dt}=\frac{\frac{1}{2}
A_{0t}^2+\sum_{n=1}^\infty\left(A_{nt}^2+B_{nt}^2\right)}
{\frac{1}{2}A_0^2+\sum_{n=1}^\infty\left(A_n^2+B_n^2\right)}+ \xi.
\end{equation}

where

\begin{eqnarray}
\label{error_Parseval}
 \xi=-\frac{\frac{1}{a}\int_{t}^{a}f(t')^2dt'}{\frac{1}{a}\int_{-a}^{a}f(t)^2dt}+ \nonumber
\\ \frac{\frac{1}{2}
\left(A_0^2-A_{0t}^2\right)+
\sum_{n=1}^\infty\left(\left(A_n^2-A_{nt}^2\right)+\left(B_n^2-B_{nt}^2\right)\right)}{\frac{1}{2}A_0^2+\sum_{n=1}^\infty\left(A_n^2+B_n^2\right)}.
\end{eqnarray}

The term $\xi$ takes into account the error produced in
Eq.\,\ref{Parseval_2} for not extending the integral limits to the
period $a$. Obviously if we set $t=a$ we obtain
$\xi=0$.\\

If we define now
\begin{equation}
\label{Rabi}
 \Omega(t)=\Omega_0e^{-2Ln2\frac{t^2}{\tau ^2}}=\Omega_0f(t),
\end{equation}

\begin{equation}
\label{Intensity}
 I(t)=I_0e^{-4Ln2\frac{t^2}{\tau ^2}}=I_0f(t)^2,
\end{equation}

\begin{equation}
\Delta=\frac{2 \pi}{\tau\epsilon},
\end{equation}

where $\epsilon$ is the number of states contained in $1/\tau$,
\begin{equation}
a=\frac{\tau\epsilon}{2}.
\end{equation}
Introducing these definitions in Eq.\,\ref{Parseval_3}, we finally
obtain

\begin{figure*}
\begin{equation}
\label{equality_demonst}
\frac{\frac{\alpha}{2}\int_{-\tau\epsilon/2}^t
\textrm{I}(t^{\prime})\textrm{d}t^{\prime}}
{\frac{\alpha}{2}\int_{-\tau\epsilon/2}^{+\tau\epsilon/2}
\textrm{I}(t^{\prime})\textrm{d}t^{\prime}}=
\frac{\sum_{n=-\infty}^{+\infty}\left[
\left[\int_{-\tau\epsilon/2}^t \frac{\Omega(t^\prime)}{2}
 \cos (n \frac{2 \pi}{\tau \epsilon}
t^\prime) dt^\prime\right]^2+\left[\int_{-\tau\epsilon/2}^t
\frac{\Omega(t^\prime)}{2} \sin (n \frac{2 \pi}{\tau \epsilon}
t^\prime)
dt^\prime\right]^2\right]}{\sum_{n=-\infty}^{+\infty}\left[\left[\int_{-\tau\epsilon/2}^{+\tau\epsilon/2}
\frac{\Omega(t^\prime)}{2} \cos (n \frac{2 \pi}{\tau \epsilon}
t^\prime)
dt^\prime\right]^2+\left[\int_{-\tau\epsilon/2}^{+\tau\epsilon/2}
\frac{\Omega(t^\prime)}{2} \sin (n \frac{2 \pi}{\tau \epsilon}
t^\prime) dt^\prime\right]^2\right]}+\xi.
\end{equation}
\end{figure*}

Equation\,\ref{equality_demonst} is the same that
Eq.\,\ref{equality} except for the error term $\xi$. This term tends
rapidly to zero when the sum is extended to a large number of
states, and we can write Eq.\,\ref{equality_demonst_2}.

\begin{widetext}
\begin{equation}
\label{equality_demonst_2}
\frac{\frac{\alpha}{2}\int_{-\tau\epsilon/2}^t
\textrm{I}(t^{\prime})\textrm{d}t^{\prime}}
{\frac{\alpha}{2}\int_{-\tau\epsilon/2}^{+\tau\epsilon/2}
\textrm{I}(t^{\prime})\textrm{d}t^{\prime}}\simeq \\
\frac{\sum_{n=-\infty}^{+\infty}\left[
\left[\int_{-\tau\epsilon/2}^t \frac{\Omega(t^\prime)}{2}
 \cos (n \frac{2 \pi}{\tau \epsilon}
t^\prime) dt^\prime\right]^2+\left[\int_{-\tau\epsilon/2}^t
\frac{\Omega(t^\prime)}{2} \sin (n \frac{2 \pi}{\tau \epsilon}
t^\prime)
dt^\prime\right]^2\right]}{\sum_{n=-\infty}^{+\infty}\left[\left[\int_{-\tau\epsilon/2}^{+\tau\epsilon/2}
\frac{\Omega(t^\prime)}{2} \cos (n \frac{2 \pi}{\tau \epsilon}
t^\prime)
dt^\prime\right]^2+\left[\int_{-\tau\epsilon/2}^{+\tau\epsilon/2}
\frac{\Omega(t^\prime)}{2} \sin (n \frac{2 \pi}{\tau \epsilon}
t^\prime) dt^\prime\right]^2\right]}.
\end{equation}
\end{widetext}
It is important to point out that in case we do not consider a large
number of states, i.e.,  $N$ does not tend to infinity, the left
side in Eq.\,\ref{Parseval_1} is not zero. Thus in this situation,
the error term $\xi$ does not only accounts for the error for not
extending the integral in Eq.\,\ref{Parseval_3} to the full period,
but also the for error produced because $S_N$ does not exactly
converge to $f(t)$.

It is instructive to analyze the behavior of the error term $\xi$ as
a function of the number of states contained in the laser bandwidth
$\epsilon$. If we choose an arbitrary point, e.g., $t=0$, and taking
into account that $\int_{0}^{a}f(t)^2dt=\int_{-a}^{a}f(t)^2dt/2$,
$A_{nt}=A_n/2$, and $B_n=0$, Eq.\,\ref{error_Parseval} can be
written as

\begin{equation}
\label{error_Parseval_simplif}
 \xi=\frac{1}{4}-\frac{\sum_{n=1}^N B_{nt}^2}{\frac{1}{2}A_0^2+\sum_{n=1}^N A_n^2}.
\end{equation}

\begin{figure}
\includegraphics[width=8cm, height=5.6cm]{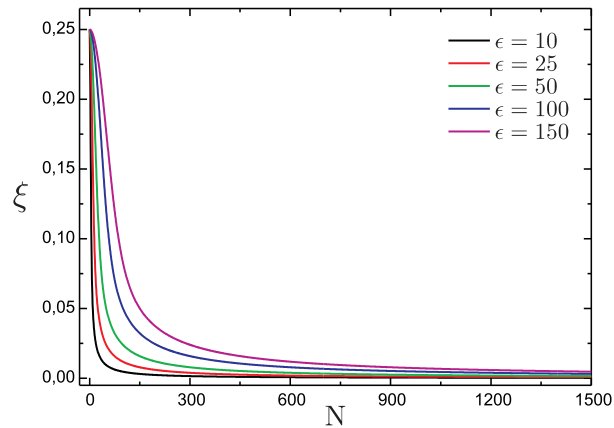}
\caption{\label{error} Error term $\xi$ as a function of N, for a
different number of states contained in the bandwidth of the pump
laser $\epsilon$ ($\tau=50\,\textrm{fs}$).}
\end{figure}

Figure\,\ref{error} shows the error term $\xi$ as a function of N
for different values of $\epsilon$. We can see that $\xi$ tends to
zero more rapidly for small values of $\epsilon$. In other words, if
the number of states inside the bandwidth is small, less states
outside the bandwidth have to be summed up to reach the incoherent
limit.

\begin{figure}
\includegraphics[width=8cm, height=5.6cm]{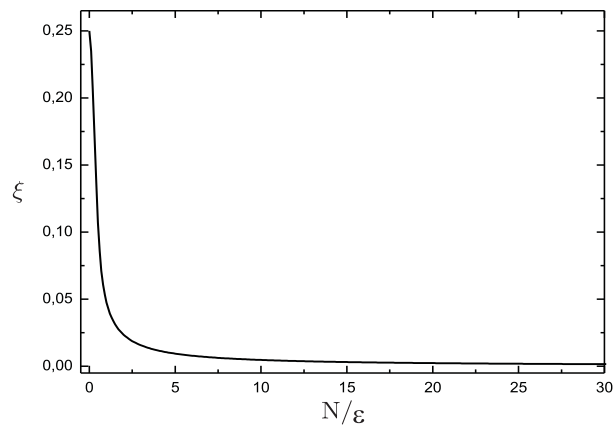}
\caption{\label{error_fullwidth} Error term $\xi$ as a function of
N/$\epsilon$.}
\end{figure}

For a better comparison Fig.\,\ref{error_fullwidth} shows $\xi$ as a
function of $N/\epsilon$, i.e., as a function of the number of
bandwidths consider in the summations. It is remarkable that for all
the different $\epsilon$ the error function takes the same value. In
other words, if one maintains constant the ratio between the number
of states inside and outside the laser bandwidth the error function
is independent of $\epsilon$. This result, as first sight non
intuitive, was already sketched in Section\,\ref{qualitative} in
terms of CPR. If we consider a situation where inside the bandwidth
there are many states, large $\epsilon$, a large number of states
lying outside the laser bandwidth will be necessary to produce the
temporal shift that yields $\xi=0$ (see Fig.\,\ref{comparison}).
According to this analysis, we can conclude that it is not necessary
to have a large density of states for ensuring the validity of the
incoherent approximation, but just an adequate balance between the
states inside and outside the pump laser bandwidth.

\section{Conclusions}

In conclusion, in the light of concepts of coherent interactions we
have demonstrated the suitability of the kinetic model to extract
the correct information from pump probe experiments in multistate
systems. The validity of the kinetic model applied to such complex
systems, has been rationalized in terms of the ratio of states
inside and outside the laser bandwidth, showing that if this
parameter is kept constant the accuracy of the kinetic model is
independent of the density of states. We believe that the field of
coherent laser matter interactions applied to problems where
historically the coherent nature of light has been sometimes
ignored, still offers a large potential for future explorations. We
should emphasize the latter results in a scenario where the constant
development of shorter pulses \cite{KRA09}, i.e., larger bandwidths,
and the increasing interest for the ultrafast dynamics of bigger
molecules \cite{OST09}, i.e., with a large density of states, demand
a correct interpretation of laser-matter interaction.

\section{Acknowledgments}

The authors thank D. Ruano for the most valuable discussions. They
also thank the Basque Government (BG) funding through a
Complementary Action and the UPV-EHU Consolidated Group Program.
Technical and human support provided by the SGIker (UPV/EHU, MICINN,
GV/EJ, ESF) is also gratefully acknowledged.

\end{document}